\documentclass[aps,prb,twocolumn,superscriptaddress,10pt]{revtex4-2}
\usepackage{graphicx} 
\usepackage{xcolor} 
\usepackage{amsmath,amssymb} 
\usepackage{bm} 
\usepackage{siunitx} 
\usepackage{booktabs} 
\usepackage{hyperref} 
\usepackage{xspace} 
\definecolor{link}{rgb}{0.1,0.1,0.9}
\definecolor{myblue}{RGB}{0,0,180}
\definecolor{myred}{RGB}{180,0,0}
\hypersetup{colorlinks=true,linkcolor=link,citecolor=link,urlcolor=link,linktocpage}
\makeatletter
\g@addto@macro\bfseries{\boldmath}
\makeatother

\begin{document}
	
\title{Magnetic Frustration in CuYbSe$_2$: an Yb-Based Triangular Lattice Selenide}
	
\author{Barkha Khattar}
\affiliation{Department of Physical Sciences, Indian Institute of Science Education and Research (IISER) Mohali, Knowledge City, Sector 81, Mohali 140306, India.}

\author{Anzar Ali}
\affiliation{Max Planck Institute for Solid State Research, Heisenbergstraße 1, D-70569 Stuttgart, Germany}
	
\author{Masahiko Isobe}
\affiliation{Max Planck Institute for Solid State Research, Heisenbergstraße 1, D-70569 Stuttgart, Germany}

\author{Yogesh Singh}
\email{yogesh@iisermohali.ac.in}
\affiliation{Department of Physical Sciences, Indian Institute of Science Education and Research (IISER) Mohali, Knowledge City, Sector 81, Mohali 140306, India.}

\begin{abstract}
		
The Yb based triangular lattice delafossites $A$Yb$X_2$ ($A$ = alkali metal, $X$ = O, S, Se) have recently been studied as quantum spin liquid candidates.  We report the synthesis of powders and single crystals of CuYbSe$_2$ with a perfect triangular lattice of Yb$^{3+}$ moments. Magnetic susceptibility and heat capacity measurements reveal no evidence of long-range magnetic ordering down to $1.8$~K in spite of a significant antiferromagnetic exchange between Yb$^{3+}$ moments, suggesting a frustrated magnetic system. Electrical resistivity measurements indicate insulating behavior, consistent with the localized nature of magnetic moments. Heat capacity reveals that CuYbSe$_2$ can be treated as an effective spin $S = 1/2$ triangular lattice antiferromagnet below $\sim 50$~K\@.  Magnetic susceptibility measurements on single crystals reveals weak magnetic anisotropy.  These properties position CuYbSe$_2$ as a promising candidate for a quantum spin liquid state and as a new platform for exploring exotic magnetic ground states in triangular lattice systems.
		
\end{abstract}  
	
\maketitle
	
\section{Introduction}
\label{Intro}
	
The study of geometrically frustrated magnetic systems remains an area of active current research, primarily due to their proclivity to host unconventional magnetic ground states \cite{Balents2010, Broholm2020, Moessne2006, Chamorro2021,Savary_2017}. In magnetic systems with spins $S = 1/2$ arranged on geometrically frustrated lattices such as triangular, kagome, or pyrochlore, conventional magnetic order is often suppressed due to the inability of all interactions to be simultaneously satisfied. Geometric frustration combined with low dimensionality are ingredients for enhanced quantum fluctuations which could lead to a highly entangled quantum spin liquid (QSL) state.  Unlike conventional magnets, this state lacks symmetry breaking magnetic order and is host to fractionalized excitations making the search for a QSL realization one of the leading areas of current research. A number of new QSL candidate materials have been reported experimentally and efforts to probe fractional excitations and entanglement properties of QSL materials continues to be developed.\cite{Chamorro2021,Savary_2017, https://doi.org/10.1002/qute.202400196,SCHEIE2025100020}

Recently rare-earth chalcogenides have been studied extensively as new platforms for exploring novel magnetic states arising from the interplay of strong spin-orbit coupling and geometrical frustration \cite{DissanayakaMudiyanselage2022,Mizutami_2022, PhysRevB.98.220409,Scheie2024,Xing2020}. Strong spin orbit coupling inherent to rare earth ions can give rise to highly anisotropic exchange interactions and help realize exotic magnetic Hamiltonians like the Kitaev model \cite{KITAEV20062, PhysRevLett.102.017205}.  In particular, materials in the family of Yb$^{3+}$ based chalcogenides, $A$YbSe$_2$ ($A =$ Li, Na, K, Cs, Ag) \cite{DissanayakaMudiyanselage2022,PhysRevB.98.220409,Scheie2024,Xing2020,Mizutami_2022, PhysRevB.103.214445, Liu_2018} have been recently investigated as candidates to explore this kind of physics. The LiYbSe$_2$ material hosts a cubic pyrochlore lattice of Yb$^{3+}$ ions which leads to geometric frustration \cite{DissanayakaMudiyanselage2022} while in AgYbSe$_2$, Yb$^{3+}$ ions form zigzag antiferromagnetic chains where the competing interactions give rise to magnetic frustration. \cite{Mizutami_2022}

In contrast to the above materials the $A$YbSe$_2$ ($A =$ Cs, K, Na) family has a quasi-two-dimensional structure and has been reported to host a perfect triangular lattice of Yb$^{3+}$ ions interacting antiferromagnetically but with no signature of long range order down to $0.4$~K, making them promising QSL candidates. \cite{PhysRevB.98.220409,Scheie2024,Xing2020}

The magnetic behavior of the $A$YbSe$_2$ triangular lattice antiferromagnets (TAFs) can be modeled using the $J_1-J_2$ model where $J_1$ and $J_2$ are nearest and next-nearest neighbor Heisenberg exchange interactions, and the $J_2/J_1$ ratio can be tuned by changing the ionic radius of the $A$ ions \cite{PhysRevB.109.014425}.  For example, the CsYbSe$_2$ (Cs$^+$ ionic radius $\approx 175$~pm) was reported to have a $J_2/J_1$ ratio which placed it in the classical $120^o$ ordered state whereas NaYbSe$_2$ (Na$^+$ ionic radius $\approx 105$~pm), with the smaller $A$ ionic size, was found to have a $J_2/J_1$ ratio which puts it proximate to the QSL state.\cite{PhysRevB.109.014425}  It was however, found that for NaYbSe$_2$, further anisotropic exchanges in addition to the Heisenberg $J_1$ and $J_2$ may be present.  Going by ionic size, Cu$^+$ ions are much smaller (Cu$^+$ ionic radius $\approx 90$~pm) than Na$^+$ ions and therefore it may be interesting to study the CuYbSe$_2$ material to see if it can be tuned completely to the QSL state predicted for the $J_1-J_2$ model in the range $0.06 \leq J_2/J_1 \leq 0.16$ \cite{PhysRevB.92.041105,PhysRevB.92.140403,PhysRevB.93.144411,PhysRevB.94.121111,PhysRevB.95.035141,PhysRevB.96.075116,PhysRevLett.123.207203}.  

The material CuYbSe$_2$ is known to exist and its crystal structure has been reported previously \cite{Daszkiewicz2008}. However, no reports of the physical or magnetic properties are available to the best of our knowledge. In this work, we have synthesized polycrystalline and single crystalline samples of CuYbSe$_2$ and report measurements of their electrical transport, anisotropic magnetic susceptibility and low temperature heat capacity.   Our study reveals that in spite of significant antiferromagnetic exchange between effective $S = 1/2$ Yb$^{3+}$ moments, there is no magnetic order or spin freezing down to $T = 1.8$~K indicating strong magnetic frustration. Additionally there is a buildup of entropy at low temperatures which is only weakly sensitive to magnetic fields.  This suggests that CuYbSe$_2$ is a new candidate in which to explore the possibility of a QSL state. 

\begin{figure}[t]
\includegraphics[width=1\linewidth]{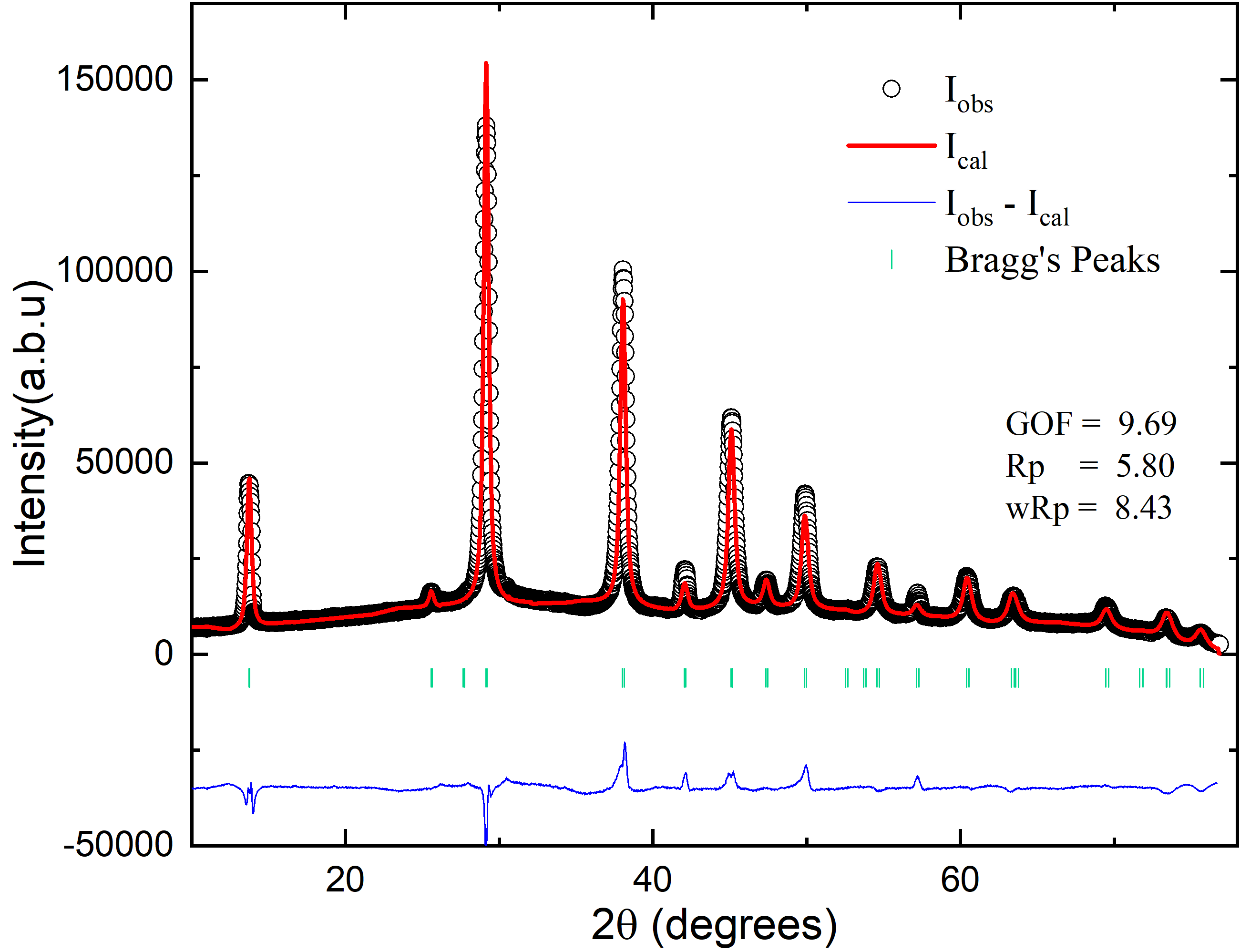}
\caption{Rietveld-refined room-temperature X-ray diffraction pattern of CuYbSe$_2$. The observed data are shown as open black circles, the calculated pattern as a solid red line, and the difference curve at the bottom as a solid blue line. Green vertical tick marks indicate the Bragg reflection positions.}
\label{Xrd}
\end{figure}

\begin{figure}
\centering
\includegraphics[width=1\linewidth]{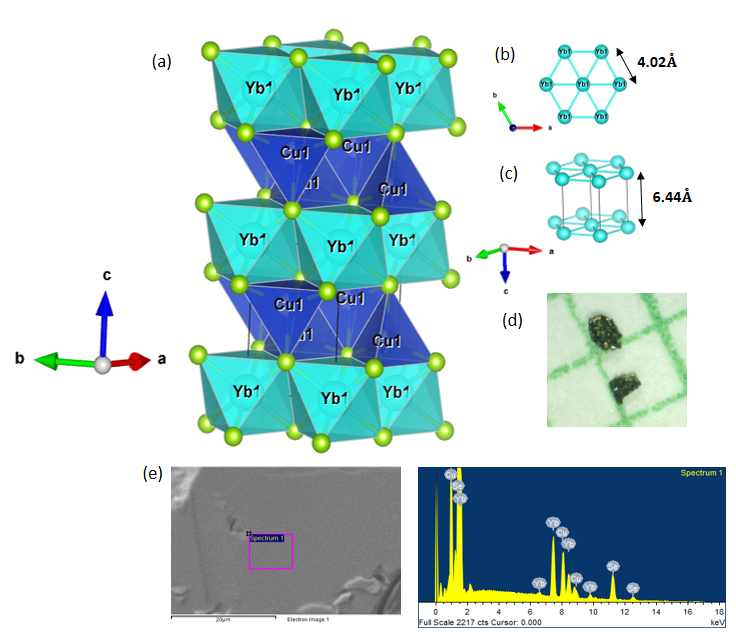}
\caption{(a) Crystal structure of CuYbSe$_2$ illustrating hexagonal stacking of Yb$^{3+}$ and Cu$^{+}$ sites in edge-sharing octahedra and tetrahedra, respectively. 
			(b) Two-dimensional triangular arrangement of Yb$^{3+}$ ions within the $ab$ plane. 
			(c) Triangular Yb layers separated by $6.44$~\AA\ along the $c$ axis, with an in-plane Yb–Yb distance of $4.02$~\AA. 
			(d) Optical micrograph of single crystals on a millimetre grid. 
			(e) Representative SEM image and EDS spectra confirming the expected Cu:Yb:Se stoichiometry.} 
\label{Structural_image}
\end{figure}

\section{Experimental Details}
\label{Exp}

Polycrystalline CuYbSe$_2$ was synthesized by conventional solid state reaction in sealed quartz ampoules. The constituent elements Cu powder (99.999\%, thermo scientific), Yb powder (99.9\%, thermo scientific) and Se shots (99.999\%, Alfa Aesar) were weighed in stochiometric ratios, thoroughly mixed, and pressed into pellets.The pellets were sealed in quartz ampoules, heated to a $1090$~\textdegree C, and held there for $12$~hr before cooling to room temperature. The single crystals were grown by heating a pellet of the same stoichiometric mixtures  at $1100$~\textdegree C for $48$~hr, cooling to $700$~\textdegree C in $10$~hr followed by quenching in air. Room-temperature X-ray diffraction (XRD) data were collected on a Rigaku SmartLab diffractometer using monochromatic Cu-K$\alpha$ radiation ($\lambda = 1.5406$~\text{\AA}), with measurements spanning $10^\circ \leq 2\theta \leq 80^\circ$ at a step size of $0.02^\circ$. Energy-dispersive X-ray (EDX) spectra were acquired using a NORAN System 7 (NSS212E) detector integrated into a Tescan Vega (TS-5130MM) scanning electron microscope. The DC magnetic susceptibility and isothermal magnetization measurements were measured using a SQUID magnetometer (Quantum Design, MPMS-XL). Heat capacity and electrical resistivity on a polycrystalline pellet were measured in a Quantum Design physical property measurement system (QD-PPMS EC-II).

\section{Results}
\label{Result}
\subsection{Crystal Structure}
CuYbSe$_2$ crystallizes in a hexagonal structure with space group ${P}$\ensuremath{\overline{3}}m1~(No.164).  A Rietveld refinement of the room-temperature powder XRD pattern (Fig.~\ref{Xrd}) confirms phase purity and yields lattice parameters given in the table~\ref{CIF}. The refined atomic positions and lattice constants are in good agreement with the reported values~\cite{Daszkiewicz2008}. 

\begin{table}
\caption{Refined structural parameters for CuYbSe$_2$ at 300~K extracted from X-ray diffraction. 
Space group \textit{P$\bar{3}$m1} (No.~164). 
Lattice parameters: $a = b = 4.0209(2)$~\AA, $c = 6.4457(3)$~\AA, $\alpha = \beta = 90^\circ$, $\gamma = 120^\circ$, $V = 90.25(1)$~\AA$^3$. Refinement residuals: $R_P = 5.79$, $wR_{p} = 8.41$, goodness-of-fit = 9.67.}
\label{CIF}
\setlength\extrarowheight{5pt}
\setlength{\tabcolsep}{5pt}
\begin{tabular}{ccccccc}
\hline
Atom & $x$ & $y$ & $z$ & Occ. & $U_{iso}$ & Site \\ \hline
Yb   & 0.00000 & 0.00000 & 0.00000 & 1.000 & 0.035 & 1$a$ \\
Cu   & 0.33333 & 0.66667 & 0.39948 & 0.500 & 0.021 & 2$d$ \\
Se   & 0.33333 & 0.66667 & 0.74793 & 1.000 & 0.020 & 2$d$ \\ \hline
\end{tabular}
\end{table}

A schematic representation of the crystal structure is shown in Fig.~\ref{Structural_image}(a). The Yb$^{3+}$ ions form ideal two-dimensional triangular layers within the $ab$-plane (Fig.~\ref{Structural_image}(b)), separated along the $c$ axis by layers of CuSe$_4$ tetrahedra (Fig.~\ref{Structural_image}(c)). This arrangement produces a quasi-two-dimensional triangular-lattice of magnetic Yb$^{3+}$ ions with an in-plane Yb–Yb separation of $4.02$~\AA\ and an interlayer spacing of $6.44$~\AA. An optical image of representative single crystals on a millimetre grid is shown in Fig.~\ref{Structural_image}(d).  Chemical composition was verified by energy-dispersive X-ray spectroscopy (EDS) on more than three crystals from the same batch. The results confirm the presence of Cu, Yb, and Se in the expected stoichiometric ratios. A scanning electron microscopy (SEM) image of a typical crystal and the corresponding EDS spectra are displayed in Fig.~\ref{Structural_image}(e). Single-crystal XRD measurements further confirmed the phase purity, space group, and lattice parameters, which are consistent with those obtained from powder diffraction and with prior reports~\cite{Daszkiewicz2008}.

\begin{figure*}
\centering
\includegraphics[width=1\linewidth]{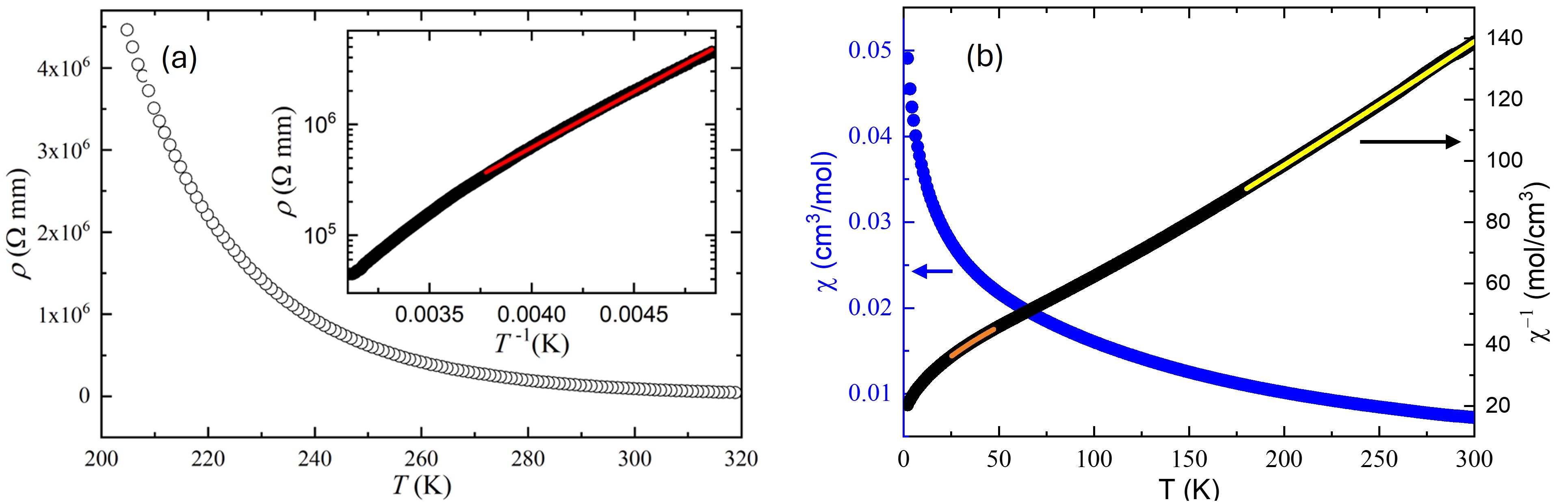}
\caption{(a) Temperature dependence of the electrical resistivity $\rho(T)$ of CuYbSe$_2$. The inset shows ln$\rho(T)$ versus 1/T, which
follows an Arrhenius behavior. (b) Magnetic susceptibility $\chi$ versus temperature T for CuYbSe$_2$ measured between $2$ and
$300$~K in a magnetic field of $\mu_0H = 0.1$~T (left axis). The inverse susceptibility 1/$\chi$ is plotted on the right axis. The solid yellow
and orange lines represent a Curie–Weiss fit to the data between $175-300$~K and $25-50$~K, respectively.}
\label{MTH}
\end{figure*}

\subsection{Resistivity}
The temperature dependence of the electrical resistivity $\rho(T)$ measured in zero field between 200 and 320~K is shown in Fig.~\ref{MTH}(a). Below 200~K, the resistivity exceeds the measurable range of our setup, indicating highly insulating behavior at low temperatures.  Within the measured range, the resistivity displays a clear thermally activated behavior that can be described by the Arrhenius relation,
\begin{equation}
	\rho(T) = \rho_0 \exp\left(\frac{E_a}{k_B T}\right),
\end{equation}
where $E_a$ is the activation energy. A linear fit to the $\ln \rho$ versus $1/T$ data [inset of Fig.~\ref{MTH}(a)] yields $E_a = 0.371$~eV, confirming thermally activated transport. These results indicate that CuYbSe$_2$ is a wide-gap semiconductor with a large resistivity and negligible conduction at low temperatures.

\subsection{Magnetic measurements}
The temperature dependence of the dc magnetic susceptibility $\chi(T)$ of CuYbSe$_2$ measured at $B = 0.1~\mathrm{T}$ is shown in Fig.~\ref{MTH}(b). At high temperatures, the susceptibility follows the Curie–Weiss law
	\begin{equation}
		\chi(T) = \chi_0 + \frac{C}{T - \theta},
	\end{equation}
\noindent where $\chi_0$ represents the temperature-independent contribution (core diamagnetism and Van Vleck paramagnetism), $C$ is the Curie constant, and $\theta$ is the Curie–Weiss temperature.

The effective magnetic moment $\mu_\mathrm{eff}$ is estimated to be $4.76~\mu_\mathrm{B}$/f.u., which is close to $4.54~\mu_\mathrm{B}$ expected for a free Yb$^{3+}$ ion, confirming the trivalent Yb state. The obtained value of $\theta = -70~\mathrm{K}$ indicates antiferromagnetic (AFM) interaction between the Yb$^{3+}$ magnetic moments. Below $20$~K,  there is a pronounced curvature in $\chi(T)$. There is however, no signature of any long range ordering down to $1.8$~K\@. Additionally, the zero-field-cooled (ZFC) and field cooled (FC) $\chi$ data at low fields $H = 100$~Oe (not shown here) did not show any cusp or sign of bifurcation ruling out any frozen magnetic state.  Heat capacity measurements, which we will present later, show that below $\sim 50$~K only the lowest Kramer's doublet is occupied and the system can be treated as an effective $S = 1/2$ system.  We therefore made a CW fit to the $\chi(T)$ data between $25$~K and $50$~K and obtained $\chi_o = 0.0100(4)$~cm$^3$/mol, $\theta = -13(1)$~K, and $C =0.81(2)$~cm$^3$ K/mol.  From this value of $C$ we estimate an effective $g$-factor of $g \approx 2.9$ assuming $S = 1/2$.

\begin{figure*}
\centering
\includegraphics[width=\linewidth]{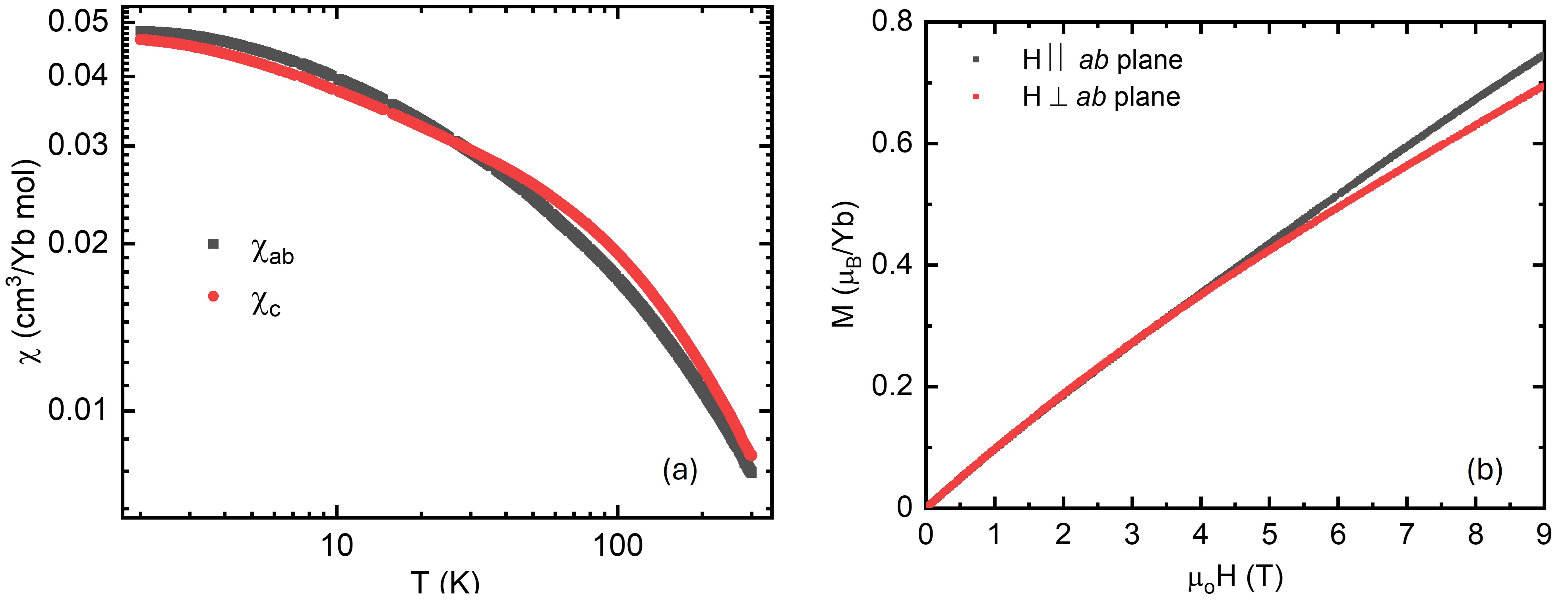}
\caption{ (a) Magnetic susceptibility measured at $H = 1$~T for fields applied parallel and perpendicular to the $ab$ plane (b) Isothermal magnetization $M(H)$ measured up to $9$~T at $T = 2$~K for fields applied parallel and perpendicular to the $ab$ plane.}
\label{Anisotropy}
\end{figure*}

The $\chi(T)$ with a magnetic field of $1$~T applied in-plane ($\chi_{ab}$) and out-of-plane ($\chi_c$), and the magnetization versus field on a single crystal of CuYbSe$_2$ are shown in Fig.~\ref{Anisotropy}(a) and (b).  There is a much weaker anisotropy than has been reported previously for alkali metal Yb delafossites $A$YbSe$_2$ ($A =$ Cs, K, Na).\cite{PhysRevB.100.224417,10.1063/5.0071161,PhysRevB.100.220407} We have fit the low temperature ($T \leq 50$~K) anisotropic $\chi(T)$ data to a CW expression and obtain the following anisotropic parameters: for $H ||$~ab-plane $\chi_0 = 1.0(1)\times 10^{-2}$~cm$^3$/mol, $\theta_{ab} = -17(3)$~K, and $g_{ab} \approx 3.6$, and for $H ||$~c-axis $\chi_0 = 1.8(4)\times 10^{-2}$~cm$^3$/mol, $\theta_c = -9(2)$~K, and $g_{c} \approx 2.7$.  Although there are bound to be larger errors in the values obtained from the fits due to the limited temperature range of data which is being fit, the overall trend of interactions ($\theta$) and effective $g$-factor being larger in-plane is the same as observed for other Yb based TAF materials \cite{PhysRevB.103.214445, Liu_2018}.  This points to the system being an XXZ magnet.  The interaction strength is larger than reported for other $A$YbSe$_2$ materials and is consistent with the Yb-Yb distance being smaller in CuYbSe$_2$.  
A similarly weak anisotropy with $M_{ab} > M_c$ is observed in the magnetization at higher fields as can be seen from Fig.~\ref{Anisotropy}(b).  The $M(H)$ doesn't saturate up to $9$~T fields and therefore an estimation of the van Vleck contribution to the magnetic susceptibility was not possible. We also note that we did not observe any $1/3$ magnetization plateaus reported for some of the other $A$YbSe$_2$ materials. 

\begin{figure*}
\centering
\includegraphics[width=1.04\linewidth]{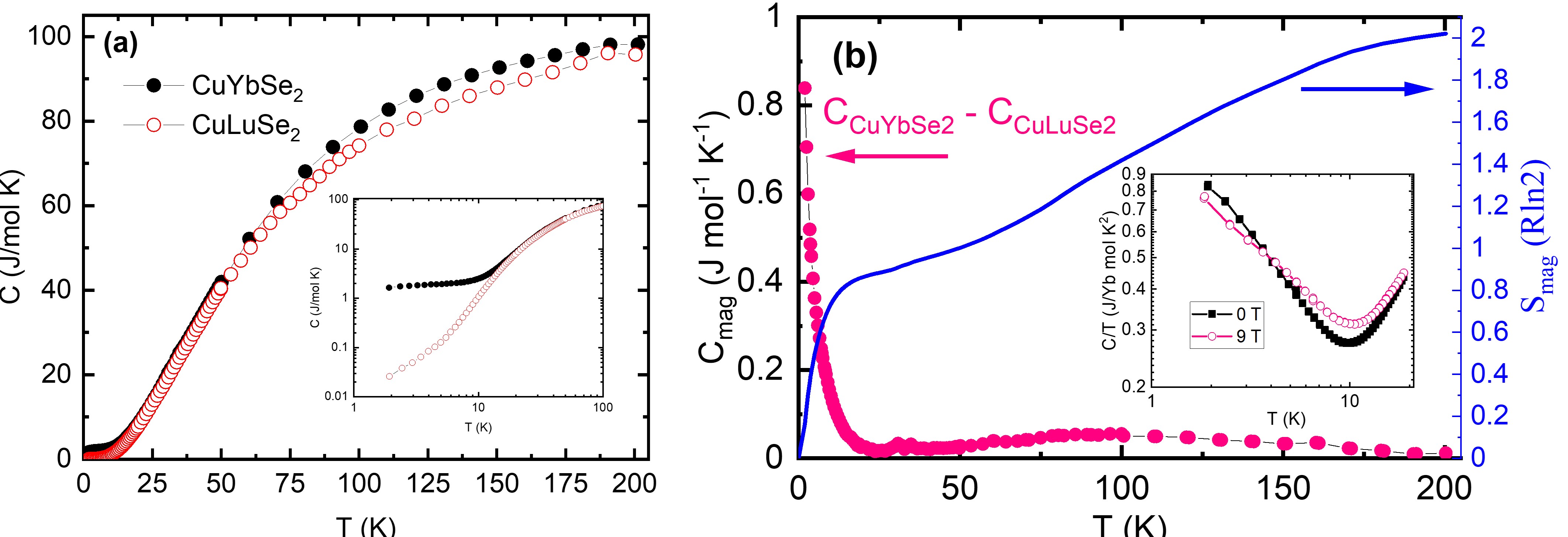}
\caption{Heat Capacity data analysis (a) Specific heat capacity $C$ plotted as a function of temperature T for CuYbSe$_2$ in black and CuLuSe$_2$ in red. The inset shows the data for both materials on a log-log scale to highlight the excess magnetic contribution below $\sim 10$~K\@.   (b) Temperature dependence of magnetic specific heat C$_{mag}$ (left-hand scale) and the corresponding magnetic entropy S$_{mag}$(right
hand scale) for CuYbSe$_2$.  The inset shows the low temperature $C/T$ data for CuYbSe$_2$ at magnetic fields of $0$ and $9$~T\@.}
	\label{HC}
\end{figure*}

\subsection{Heat Capacity}
In insulating magnetic materials, the total heat capacity ($C$) consists of contributions from phononic ($C_{ph}$) and
magnetic (C$_{mag}$) degrees of freedom. Ideally, these components can be separated by measuring a nonmagnetic structural analog. We have measured the heat capacity of isostructural CuLuSe$_2$ and will use it as the phonon contribution to the heat capacity of CuYbSe$_2$.
The data of zero field specific heat $C$ for both are shown as a function of temperature on a linear scale in Fig.~\ref{HC}(a). No sharp anomaly indicative of a long ranged magnetic order is observed down to $1.8$~K\@.

Excess magnetic contribution to the heat capacity can be seen below $\sim 10$~K in Fig.~\ref{HC}(a) inset where the $C$ versus $T$ for both materials is shown on a log-log scale.  This low temperature magnetic contribution in CuYbSe$_2$ is similar to that reported for other $A$YbSe$_2$ materials and seems to be universal for Yb based TAFs.  

The magnetic ground state can also be determined from the low temperature behavior of the magnetic heat capacity and entropy.  The magnetic heat capacity $C_\mathrm{mag}(T)$ of CuYbSe$_2$, obtained by subtracting the scaled $C(T)$ of CuLuSe$_2$ from that of CuYbSe$_2$, is shown as $C_\mathrm{mag}/T$ in Fig.~\ref{HC}(b) (left axis). To account for the differences in formula masses between CuYbSe$_2$ and the CuLuSe$_2$ compound, the lattice contribution from CuLuSe$_2$ was adjusted by rescaling the temperature, $T$, according to the relation:

\begin{equation}
	T^* = \frac{T}{\left( \frac{{\rm M}_{\mathrm{CuYbSe}_2}}{{\rm M}_{\mathrm{CuLuSe}_2}} \right)^{1/2}},
	\label{eq1}
\end{equation}

\noindent as derived from the Debye model, where the Debye temperature $\theta_{\text{D}}$ scales with the inverse square root of the formula mass, $\theta_{\text{D}} \sim M^{-1/2}$ or $T/\theta_{\text{D}} \sim M^{1/2}$ \cite{Kittel2004}.

The $C_{mag}/T$ shows a broad hump around $100$~K, most likely from the excitation to the higher CEF levels.  The $C_{mag}$ shows a sharp increase below $T \sim 10$~K\@.  We calculated $S_{mag}$ by integrating the $C_{mag}/T$ data with respect to 
the temperature. The entropy increases sharply on increasing temperature from $1.8$~K and shows a region below $50$~K where $S_{mag}$ saturates at $\approx 90\% {\rm R}ln2$, indicating that below $50$~K the lowest Kramer's doublet is occupied.  This strongly supports the treatment of CuYbSe$_2$ as an effective $S = 1/2$ TAF at low temperatures.  The inset in Fig.~\ref{HC}(b) shows the $C/T$ data for CuYbSe$_2$ at $0$ and $9$~T applied fields.  No dramatic change is observed.  This shows that the $C_{mag}$ is only weakly sensitive to magnetic fields.  In particular, we did not observe any field induced magnetic ordering as has been reported for some $A$YbSe$_2$ materials \cite{PhysRevB.98.220409,Scheie2024,Xing2020}.

\subsection{Conclusion}
We have synthesized polycrystal and single crystal samples of CuYbSe$_2$ and studied their magnetic properties.  The material crystallizes in the hexagonal structure with space group ${P}$\ensuremath{\overline{3}}m1~(No.164) with Yb$^{3+}$ ions forming a perfect triangular lattice.  The Yb-Yb in-plane distance ($4.02$\AA) is shorter than found in other $A$YbSe$_2$ materials reported before and therefore leads to stronger in-plane exchange between the Yb moments.  Additionally, according to the reported phase diagram, the lattice parameters of CuYbSe$_2$ suggest that it should lie closer to the quantum spin liquid regime expected in the $J_1-J_2$ model \cite{PhysRevB.109.014425}.  Our magnetic susceptibility and heat capacity measurements confirm that at high temperatures we have Yb$^{3+}$ local moments which interact antiferromagnetically.  At low temperature ($\leq 50$~K) CEF effects lead to a Kramer's doublet ground state with an effective $S = 1/2$.  Thus, at low temperature the system can be viewed as a triangular lattice antiferromagnet (TAF) with effective $S = 1/2$ moments.  From magnetic measurements on single crystals we find weak magnetic anisotropy with $\theta_{ab} = -17(3)$~K, and $g_{ab} \approx 3.6$, and $\theta_c = -9(2)$~K, and $g_{c} \approx 2.7$ at low temperatures.  In spite of the significant exchange there is no signature of magnetic order down to $1.8$~K indicating that strong magnetic frustration is at play.   This is consistent with the thermodynamic properties of other Yb based triangular lattice chalcogenides like NaYbS$_2$ and NaYbSe$_2$ \cite{Liu_2018,PhysRevB.100.224417}.  There is however one main point of difference.  While these other materials show a large anisotropy in their magnetic properties (e.g. for NaYbSe$_2$, $g_\perp = 3.13$ and $g_{||}= 1$ and for NaYbS$_2$, $g_\perp = 3.19$ and $g_{||}= 0.57$), there is very weak magnetic anisotropy for CuYbSe$_2$.  Measurements to lower temperatures are desirable to confirm a QSL state. Additionally, it would be interesting to do $M(H)$ measurements to higher fields to look for $1/3$ plateaus which have been reported for some of the other Yb based TAFs.  Finally a direct estimation of the exchanges $J_1$ and $J_2$ from neutron measurements will help place it on the phase diagram.

\section*{Acknowledgement}
We thank the X-ray, the SEM, and the SQUID magnetometer facilities at IISER Mohali and Max planck Institute for Solid state research, Stuttgart.  BK acknowledges IISER Mohali for institute fellowship.  Y. S. acknowledges support from SERB project CRG/2022/000015 and STARS project STARS-1/240.

\bibliography{Ref}

\begin{thebibliography}{29}%
\makeatletter
\providecommand \@ifxundefined [1]{%
 \@ifx{#1\undefined}
}%
\providecommand \@ifnum [1]{%
 \ifnum #1\expandafter \@firstoftwo
 \else \expandafter \@secondoftwo
 \fi
}%
\providecommand \@ifx [1]{%
 \ifx #1\expandafter \@firstoftwo
 \else \expandafter \@secondoftwo
 \fi
}%
\providecommand \natexlab [1]{#1}%
\providecommand \enquote  [1]{``#1''}%
\providecommand \bibnamefont  [1]{#1}%
\providecommand \bibfnamefont [1]{#1}%
\providecommand \citenamefont [1]{#1}%
\providecommand \href@noop [0]{\@secondoftwo}%
\providecommand \href [0]{\begingroup \@sanitize@url \@href}%
\providecommand \@href[1]{\@@startlink{#1}\@@href}%
\providecommand \@@href[1]{\endgroup#1\@@endlink}%
\providecommand \@sanitize@url [0]{\catcode `\\12\catcode `\$12\catcode `\&12\catcode `\#12\catcode `\^12\catcode `\_12\catcode `\%12\relax}%
\providecommand \@@startlink[1]{}%
\providecommand \@@endlink[0]{}%
\providecommand \url  [0]{\begingroup\@sanitize@url \@url }%
\providecommand \@url [1]{\endgroup\@href {#1}{\urlprefix }}%
\providecommand \urlprefix  [0]{URL }%
\providecommand \Eprint [0]{\href }%
\providecommand \doibase [0]{https://doi.org/}%
\providecommand \selectlanguage [0]{\@gobble}%
\providecommand \bibinfo  [0]{\@secondoftwo}%
\providecommand \bibfield  [0]{\@secondoftwo}%
\providecommand \translation [1]{[#1]}%
\providecommand \BibitemOpen [0]{}%
\providecommand \bibitemStop [0]{}%
\providecommand \bibitemNoStop [0]{.\EOS\space}%
\providecommand \EOS [0]{\spacefactor3000\relax}%
\providecommand \BibitemShut  [1]{\csname bibitem#1\endcsname}%
\let\auto@bib@innerbib\@empty
\bibitem [{\citenamefont {Balents}(2010)}]{Balents2010}%
  \BibitemOpen
  \bibfield  {author} {\bibinfo {author} {\bibfnamefont {L.}~\bibnamefont {Balents}},\ }\bibfield  {title} {\bibinfo {title} {Spin liquids in frustrated magnets},\ }\href {https://doi.org/10.1038/nature08917} {\bibfield  {journal} {\bibinfo  {journal} {Nature}\ }\textbf {\bibinfo {volume} {464}},\ \bibinfo {pages} {199} (\bibinfo {year} {2010})}\BibitemShut {NoStop}%
\bibitem [{\citenamefont {Broholm}\ \emph {et~al.}(2020)\citenamefont {Broholm}, \citenamefont {Cava}, \citenamefont {Kivelson}, \citenamefont {Nocera}, \citenamefont {Norman},\ and\ \citenamefont {Senthil}}]{Broholm2020}%
  \BibitemOpen
  \bibfield  {author} {\bibinfo {author} {\bibfnamefont {C.}~\bibnamefont {Broholm}}, \bibinfo {author} {\bibfnamefont {R.~J.}\ \bibnamefont {Cava}}, \bibinfo {author} {\bibfnamefont {S.~A.}\ \bibnamefont {Kivelson}}, \bibinfo {author} {\bibfnamefont {D.~G.}\ \bibnamefont {Nocera}}, \bibinfo {author} {\bibfnamefont {M.~R.}\ \bibnamefont {Norman}},\ and\ \bibinfo {author} {\bibfnamefont {T.}~\bibnamefont {Senthil}},\ }\bibfield  {title} {\bibinfo {title} {Quantum spin liquids},\ }\href {https://doi.org/10.1126/science.aay0668} {\bibfield  {journal} {\bibinfo  {journal} {Science}\ }\textbf {\bibinfo {volume} {367}},\ \bibinfo {pages} {eaay0668} (\bibinfo {year} {2020})}\BibitemShut {NoStop}%
\bibitem [{\citenamefont {Moessner}\ and\ \citenamefont {Ramirez}(2006)}]{Moessne2006}%
  \BibitemOpen
  \bibfield  {author} {\bibinfo {author} {\bibfnamefont {R.}~\bibnamefont {Moessner}}\ and\ \bibinfo {author} {\bibfnamefont {A.~P.}\ \bibnamefont {Ramirez}},\ }\bibfield  {title} {\bibinfo {title} {Geometrical frustration},\ }\href {https://doi.org/10.1063/1.2186278} {\bibfield  {journal} {\bibinfo  {journal} {Physics Today}\ }\textbf {\bibinfo {volume} {59}},\ \bibinfo {pages} {24} (\bibinfo {year} {2006})}\BibitemShut {NoStop}%
\bibitem [{\citenamefont {Chamorro}\ \emph {et~al.}(2021)\citenamefont {Chamorro}, \citenamefont {McQueen},\ and\ \citenamefont {Tran}}]{Chamorro2021}%
  \BibitemOpen
  \bibfield  {author} {\bibinfo {author} {\bibfnamefont {J.~R.}\ \bibnamefont {Chamorro}}, \bibinfo {author} {\bibfnamefont {T.~M.}\ \bibnamefont {McQueen}},\ and\ \bibinfo {author} {\bibfnamefont {T.~T.}\ \bibnamefont {Tran}},\ }\bibfield  {title} {\bibinfo {title} {Chemistry of {Quantum Spin Liquids}},\ }\href {https://doi.org/10.1021/acs.chemrev.0c00641} {\bibfield  {journal} {\bibinfo  {journal} {Chemical Reviews}\ }\textbf {\bibinfo {volume} {121}},\ \bibinfo {pages} {2898} (\bibinfo {year} {2021})}\BibitemShut {NoStop}%
\bibitem [{\citenamefont {Savary}\ and\ \citenamefont {Balents}(2016)}]{Savary_2017}%
  \BibitemOpen
  \bibfield  {author} {\bibinfo {author} {\bibfnamefont {L.}~\bibnamefont {Savary}}\ and\ \bibinfo {author} {\bibfnamefont {L.}~\bibnamefont {Balents}},\ }\bibfield  {title} {\bibinfo {title} {Quantum spin liquids: a review},\ }\href {https://doi.org/10.1088/0034-4885/80/1/016502} {\bibfield  {journal} {\bibinfo  {journal} {Reports on Progress in Physics}\ }\textbf {\bibinfo {volume} {80}},\ \bibinfo {pages} {016502} (\bibinfo {year} {2016})}\BibitemShut {NoStop}%
\bibitem [{\citenamefont {Laurell}\ \emph {et~al.}(2025)\citenamefont {Laurell}, \citenamefont {Scheie}, \citenamefont {Dagotto},\ and\ \citenamefont {Tennant}}]{https://doi.org/10.1002/qute.202400196}%
  \BibitemOpen
  \bibfield  {author} {\bibinfo {author} {\bibfnamefont {P.}~\bibnamefont {Laurell}}, \bibinfo {author} {\bibfnamefont {A.}~\bibnamefont {Scheie}}, \bibinfo {author} {\bibfnamefont {E.}~\bibnamefont {Dagotto}},\ and\ \bibinfo {author} {\bibfnamefont {D.~A.}\ \bibnamefont {Tennant}},\ }\bibfield  {title} {\bibinfo {title} {{Witnessing Entanglement and Quantum Correlations in Condensed Matter: A Review}},\ }\href {https://doi.org/https://doi.org/10.1002/qute.202400196} {\bibfield  {journal} {\bibinfo  {journal} {Advanced Quantum Technologies}\ }\textbf {\bibinfo {volume} {8}},\ \bibinfo {pages} {2400196} (\bibinfo {year} {2025})}\BibitemShut {NoStop}%
\bibitem [{\citenamefont {Scheie}\ \emph {et~al.}(2025)\citenamefont {Scheie}, \citenamefont {Laurell}, \citenamefont {Simeth}, \citenamefont {Dagotto},\ and\ \citenamefont {Tennant}}]{SCHEIE2025100020}%
  \BibitemOpen
  \bibfield  {author} {\bibinfo {author} {\bibfnamefont {A.}~\bibnamefont {Scheie}}, \bibinfo {author} {\bibfnamefont {P.}~\bibnamefont {Laurell}}, \bibinfo {author} {\bibfnamefont {W.}~\bibnamefont {Simeth}}, \bibinfo {author} {\bibfnamefont {E.}~\bibnamefont {Dagotto}},\ and\ \bibinfo {author} {\bibfnamefont {D.~A.}\ \bibnamefont {Tennant}},\ }\bibfield  {title} {\bibinfo {title} {Tutorial: Extracting entanglement signatures from neutron spectroscopy},\ }\href {https://doi.org/https://doi.org/10.1016/j.mtquan.2024.100020} {\bibfield  {journal} {\bibinfo  {journal} {Materials Today Quantum}\ }\textbf {\bibinfo {volume} {5}},\ \bibinfo {pages} {100020} (\bibinfo {year} {2025})}\BibitemShut {NoStop}%
\bibitem [{\citenamefont {Dissanayaka~Mudiyanselage}\ \emph {et~al.}(2022)\citenamefont {Dissanayaka~Mudiyanselage}, \citenamefont {Wang}, \citenamefont {Vilella}, \citenamefont {Mourigal}, \citenamefont {Kotliar},\ and\ \citenamefont {Xie}}]{DissanayakaMudiyanselage2022}%
  \BibitemOpen
  \bibfield  {author} {\bibinfo {author} {\bibfnamefont {R.~S.}\ \bibnamefont {Dissanayaka~Mudiyanselage}}, \bibinfo {author} {\bibfnamefont {H.}~\bibnamefont {Wang}}, \bibinfo {author} {\bibfnamefont {O.}~\bibnamefont {Vilella}}, \bibinfo {author} {\bibfnamefont {M.}~\bibnamefont {Mourigal}}, \bibinfo {author} {\bibfnamefont {G.}~\bibnamefont {Kotliar}},\ and\ \bibinfo {author} {\bibfnamefont {W.}~\bibnamefont {Xie}},\ }\bibfield  {title} {\bibinfo {title} {{LiYbSe$_2$}: Frustrated magnetism in the pyrochlore lattice},\ }\href {https://doi.org/10.1021/jacs.2c02839} {\bibfield  {journal} {\bibinfo  {journal} {Journal of the American Chemical Society}\ }\textbf {\bibinfo {volume} {144}},\ \bibinfo {pages} {11933} (\bibinfo {year} {2022})}\BibitemShut {NoStop}%
\bibitem [{\citenamefont {Mizutami}\ \emph {et~al.}(2022)\citenamefont {Mizutami}, \citenamefont {Ohmagari}, \citenamefont {Onimaru}, \citenamefont {Shimura}, \citenamefont {Yamamoto}, \citenamefont {Umeo},\ and\ \citenamefont {Takabatake}}]{Mizutami_2022}%
  \BibitemOpen
  \bibfield  {author} {\bibinfo {author} {\bibfnamefont {S.}~\bibnamefont {Mizutami}}, \bibinfo {author} {\bibfnamefont {Y.}~\bibnamefont {Ohmagari}}, \bibinfo {author} {\bibfnamefont {T.}~\bibnamefont {Onimaru}}, \bibinfo {author} {\bibfnamefont {Y.}~\bibnamefont {Shimura}}, \bibinfo {author} {\bibfnamefont {R.}~\bibnamefont {Yamamoto}}, \bibinfo {author} {\bibfnamefont {K.}~\bibnamefont {Umeo}},\ and\ \bibinfo {author} {\bibfnamefont {T.}~\bibnamefont {Takabatake}},\ }\bibfield  {title} {\bibinfo {title} {Magnetic properties of rare-earth zigzag chain systems {RAgSe$_2$} ({R} = {H}o, {E}r, {T}m, and {Y}b)},\ }\href {https://doi.org/10.1088/1742-6596/2164/1/012025} {\bibfield  {journal} {\bibinfo  {journal} {Journal of Physics: Conference Series}\ }\textbf {\bibinfo {volume} {2164}},\ \bibinfo {pages} {012025} (\bibinfo {year} {2022})}\BibitemShut {NoStop}%
\bibitem [{\citenamefont {Baenitz}\ \emph {et~al.}(2018)\citenamefont {Baenitz}, \citenamefont {Schlender}, \citenamefont {Sichelschmidt}, \citenamefont {Onykiienko}, \citenamefont {Zangeneh}, \citenamefont {Ranjith}, \citenamefont {Sarkar}, \citenamefont {Hozoi}, \citenamefont {Walker}, \citenamefont {Orain}, \citenamefont {Yasuoka}, \citenamefont {van~den Brink}, \citenamefont {Klauss}, \citenamefont {Inosov},\ and\ \citenamefont {Doert}}]{PhysRevB.98.220409}%
  \BibitemOpen
  \bibfield  {author} {\bibinfo {author} {\bibfnamefont {M.}~\bibnamefont {Baenitz}}, \bibinfo {author} {\bibfnamefont {P.}~\bibnamefont {Schlender}}, \bibinfo {author} {\bibfnamefont {J.}~\bibnamefont {Sichelschmidt}}, \bibinfo {author} {\bibfnamefont {Y.~A.}\ \bibnamefont {Onykiienko}}, \bibinfo {author} {\bibfnamefont {Z.}~\bibnamefont {Zangeneh}}, \bibinfo {author} {\bibfnamefont {K.~M.}\ \bibnamefont {Ranjith}}, \bibinfo {author} {\bibfnamefont {R.}~\bibnamefont {Sarkar}}, \bibinfo {author} {\bibfnamefont {L.}~\bibnamefont {Hozoi}}, \bibinfo {author} {\bibfnamefont {H.~C.}\ \bibnamefont {Walker}}, \bibinfo {author} {\bibfnamefont {J.-C.}\ \bibnamefont {Orain}}, \bibinfo {author} {\bibfnamefont {H.}~\bibnamefont {Yasuoka}}, \bibinfo {author} {\bibfnamefont {J.}~\bibnamefont {van~den Brink}}, \bibinfo {author} {\bibfnamefont {H.~H.}\ \bibnamefont {Klauss}}, \bibinfo {author} {\bibfnamefont {D.~S.}\ \bibnamefont {Inosov}},\ and\ \bibinfo {author} {\bibfnamefont {T.}~\bibnamefont {Doert}},\ }\bibfield
  {title} {\bibinfo {title} {{NaYbSe$_2$}: A planar spin-$\frac{1}{2}$ triangular-lattice magnet and putative spin liquid},\ }\href {https://doi.org/10.1103/PhysRevB.98.220409} {\bibfield  {journal} {\bibinfo  {journal} {Phys. Rev. B}\ }\textbf {\bibinfo {volume} {98}},\ \bibinfo {pages} {220409} (\bibinfo {year} {2018})}\BibitemShut {NoStop}%
\bibitem [{\citenamefont {Scheie}\ \emph {et~al.}(2024{\natexlab{a}})\citenamefont {Scheie}, \citenamefont {Ghioldi}, \citenamefont {Xing}, \citenamefont {Paddison}, \citenamefont {Sherman}, \citenamefont {Dupont}, \citenamefont {Sanjeewa}, \citenamefont {Lee}, \citenamefont {Woods}, \citenamefont {Abernathy}, \citenamefont {Pajerowski}, \citenamefont {Williams}, \citenamefont {Zhang}, \citenamefont {Manuel}, \citenamefont {Trumper}, \citenamefont {Pemmaraju}, \citenamefont {Sefat}, \citenamefont {Parker}, \citenamefont {Devereaux}, \citenamefont {Movshovich}, \citenamefont {Moore}, \citenamefont {Batista},\ and\ \citenamefont {Tennant}}]{Scheie2024}%
  \BibitemOpen
  \bibfield  {author} {\bibinfo {author} {\bibfnamefont {A.~O.}\ \bibnamefont {Scheie}}, \bibinfo {author} {\bibfnamefont {E.~A.}\ \bibnamefont {Ghioldi}}, \bibinfo {author} {\bibfnamefont {J.}~\bibnamefont {Xing}}, \bibinfo {author} {\bibfnamefont {J.~A.~M.}\ \bibnamefont {Paddison}}, \bibinfo {author} {\bibfnamefont {N.~E.}\ \bibnamefont {Sherman}}, \bibinfo {author} {\bibfnamefont {M.}~\bibnamefont {Dupont}}, \bibinfo {author} {\bibfnamefont {L.~D.}\ \bibnamefont {Sanjeewa}}, \bibinfo {author} {\bibfnamefont {S.}~\bibnamefont {Lee}}, \bibinfo {author} {\bibfnamefont {A.~J.}\ \bibnamefont {Woods}}, \bibinfo {author} {\bibfnamefont {D.}~\bibnamefont {Abernathy}}, \bibinfo {author} {\bibfnamefont {D.~M.}\ \bibnamefont {Pajerowski}}, \bibinfo {author} {\bibfnamefont {T.~J.}\ \bibnamefont {Williams}}, \bibinfo {author} {\bibfnamefont {S.-S.}\ \bibnamefont {Zhang}}, \bibinfo {author} {\bibfnamefont {L.~O.}\ \bibnamefont {Manuel}}, \bibinfo {author} {\bibfnamefont {A.~E.}\ \bibnamefont {Trumper}}, \bibinfo
  {author} {\bibfnamefont {C.~D.}\ \bibnamefont {Pemmaraju}}, \bibinfo {author} {\bibfnamefont {A.~S.}\ \bibnamefont {Sefat}}, \bibinfo {author} {\bibfnamefont {D.~S.}\ \bibnamefont {Parker}}, \bibinfo {author} {\bibfnamefont {T.~P.}\ \bibnamefont {Devereaux}}, \bibinfo {author} {\bibfnamefont {R.}~\bibnamefont {Movshovich}}, \bibinfo {author} {\bibfnamefont {J.~E.}\ \bibnamefont {Moore}}, \bibinfo {author} {\bibfnamefont {C.~D.}\ \bibnamefont {Batista}},\ and\ \bibinfo {author} {\bibfnamefont {D.~A.}\ \bibnamefont {Tennant}},\ }\bibfield  {title} {\bibinfo {title} {Proximate spin liquid and fractionalization in the triangular antiferromagnet {KYbSe$_2$}},\ }\href {https://doi.org/10.1038/s41567-023-02259-1} {\bibfield  {journal} {\bibinfo  {journal} {Nature Physics}\ }\textbf {\bibinfo {volume} {20}},\ \bibinfo {pages} {74} (\bibinfo {year} {2024}{\natexlab{a}})}\BibitemShut {NoStop}%
\bibitem [{\citenamefont {Xing}\ \emph {et~al.}(2020)\citenamefont {Xing}, \citenamefont {Sanjeewa}, \citenamefont {Kim}, \citenamefont {Stewart}, \citenamefont {Du}, \citenamefont {Reboredo}, \citenamefont {Custelcean},\ and\ \citenamefont {Sefat}}]{Xing2020}%
  \BibitemOpen
  \bibfield  {author} {\bibinfo {author} {\bibfnamefont {J.}~\bibnamefont {Xing}}, \bibinfo {author} {\bibfnamefont {L.~D.}\ \bibnamefont {Sanjeewa}}, \bibinfo {author} {\bibfnamefont {J.}~\bibnamefont {Kim}}, \bibinfo {author} {\bibfnamefont {G.~R.}\ \bibnamefont {Stewart}}, \bibinfo {author} {\bibfnamefont {M.-H.}\ \bibnamefont {Du}}, \bibinfo {author} {\bibfnamefont {F.~A.}\ \bibnamefont {Reboredo}}, \bibinfo {author} {\bibfnamefont {R.}~\bibnamefont {Custelcean}},\ and\ \bibinfo {author} {\bibfnamefont {A.~S.}\ \bibnamefont {Sefat}},\ }\bibfield  {title} {\bibinfo {title} {Crystal synthesis and frustrated magnetism in triangular lattice {CsRESe$_2$} ({RE} = {La}--{Lu}): Quantum spin liquid candidates {CsCeSe$_2$} and {CsYbSe$_2$}},\ }\href {https://doi.org/10.1021/acsmaterialslett.9b00464} {\bibfield  {journal} {\bibinfo  {journal} {ACS Materials Letters}\ }\textbf {\bibinfo {volume} {2}},\ \bibinfo {pages} {71} (\bibinfo {year} {2020})}\BibitemShut {NoStop}%
\bibitem [{\citenamefont {Kitaev}(2006)}]{KITAEV20062}%
  \BibitemOpen
  \bibfield  {author} {\bibinfo {author} {\bibfnamefont {A.}~\bibnamefont {Kitaev}},\ }\bibfield  {title} {\bibinfo {title} {Anyons in an exactly solved model and beyond},\ }\href {https://doi.org/https://doi.org/10.1016/j.aop.2005.10.005} {\bibfield  {journal} {\bibinfo  {journal} {Annals of Physics}\ }\textbf {\bibinfo {volume} {321}},\ \bibinfo {pages} {2} (\bibinfo {year} {2006})},\ \bibinfo {note} {{{J}anuary Special Issue}}\BibitemShut {NoStop}%
\bibitem [{\citenamefont {Jackeli}\ and\ \citenamefont {Khaliullin}(2009)}]{PhysRevLett.102.017205}%
  \BibitemOpen
  \bibfield  {author} {\bibinfo {author} {\bibfnamefont {G.}~\bibnamefont {Jackeli}}\ and\ \bibinfo {author} {\bibfnamefont {G.}~\bibnamefont {Khaliullin}},\ }\bibfield  {title} {\bibinfo {title} {Mott insulators in the strong spin-orbit coupling limit: From heisenberg to a quantum compass and kitaev models},\ }\href {https://doi.org/10.1103/PhysRevLett.102.017205} {\bibfield  {journal} {\bibinfo  {journal} {Phys. Rev. Lett.}\ }\textbf {\bibinfo {volume} {102}},\ \bibinfo {pages} {017205} (\bibinfo {year} {2009})}\BibitemShut {NoStop}%
\bibitem [{\citenamefont {Schmidt}\ \emph {et~al.}(2021)\citenamefont {Schmidt}, \citenamefont {Sichelschmidt}, \citenamefont {Ranjith}, \citenamefont {Doert},\ and\ \citenamefont {Baenitz}}]{PhysRevB.103.214445}%
  \BibitemOpen
  \bibfield  {author} {\bibinfo {author} {\bibfnamefont {B.}~\bibnamefont {Schmidt}}, \bibinfo {author} {\bibfnamefont {J.}~\bibnamefont {Sichelschmidt}}, \bibinfo {author} {\bibfnamefont {K.~M.}\ \bibnamefont {Ranjith}}, \bibinfo {author} {\bibfnamefont {T.}~\bibnamefont {Doert}},\ and\ \bibinfo {author} {\bibfnamefont {M.}~\bibnamefont {Baenitz}},\ }\bibfield  {title} {\bibinfo {title} {Yb delafossites: Unique exchange frustration of $4f$ spin-$\frac{1}{2}$ moments on a perfect triangular lattice},\ }\href {https://doi.org/10.1103/PhysRevB.103.214445} {\bibfield  {journal} {\bibinfo  {journal} {Phys. Rev. B}\ }\textbf {\bibinfo {volume} {103}},\ \bibinfo {pages} {214445} (\bibinfo {year} {2021})}\BibitemShut {NoStop}%
\bibitem [{\citenamefont {Liu}\ \emph {et~al.}(2018)\citenamefont {Liu}, \citenamefont {Zhang}, \citenamefont {Ji}, \citenamefont {Liu}, \citenamefont {Li}, \citenamefont {Wang}, \citenamefont {Lei}, \citenamefont {Chen},\ and\ \citenamefont {Zhang}}]{Liu_2018}%
  \BibitemOpen
  \bibfield  {author} {\bibinfo {author} {\bibfnamefont {W.}~\bibnamefont {Liu}}, \bibinfo {author} {\bibfnamefont {Z.}~\bibnamefont {Zhang}}, \bibinfo {author} {\bibfnamefont {J.}~\bibnamefont {Ji}}, \bibinfo {author} {\bibfnamefont {Y.}~\bibnamefont {Liu}}, \bibinfo {author} {\bibfnamefont {J.}~\bibnamefont {Li}}, \bibinfo {author} {\bibfnamefont {X.}~\bibnamefont {Wang}}, \bibinfo {author} {\bibfnamefont {H.}~\bibnamefont {Lei}}, \bibinfo {author} {\bibfnamefont {G.}~\bibnamefont {Chen}},\ and\ \bibinfo {author} {\bibfnamefont {Q.}~\bibnamefont {Zhang}},\ }\bibfield  {title} {\bibinfo {title} {Rare-earth chalcogenides: A large family of triangular lattice spin liquid candidates},\ }\href {https://doi.org/10.1088/0256-307x/35/11/117501} {\bibfield  {journal} {\bibinfo  {journal} {Chinese Physics Letters}\ }\textbf {\bibinfo {volume} {35}},\ \bibinfo {pages} {117501} (\bibinfo {year} {2018})}\BibitemShut {NoStop}%
\bibitem [{\citenamefont {Scheie}\ \emph {et~al.}(2024{\natexlab{b}})\citenamefont {Scheie}, \citenamefont {Kamiya}, \citenamefont {Zhang}, \citenamefont {Lee}, \citenamefont {Woods}, \citenamefont {Ajeesh}, \citenamefont {Gonzalez}, \citenamefont {Bernu}, \citenamefont {Villanova}, \citenamefont {Xing}, \citenamefont {Huang}, \citenamefont {Zhang}, \citenamefont {Ma}, \citenamefont {Choi}, \citenamefont {Pajerowski}, \citenamefont {Zhou}, \citenamefont {Sefat}, \citenamefont {Okamoto}, \citenamefont {Berlijn}, \citenamefont {Messio}, \citenamefont {Movshovich}, \citenamefont {Batista},\ and\ \citenamefont {Tennant}}]{PhysRevB.109.014425}%
  \BibitemOpen
  \bibfield  {author} {\bibinfo {author} {\bibfnamefont {A.~O.}\ \bibnamefont {Scheie}}, \bibinfo {author} {\bibfnamefont {Y.}~\bibnamefont {Kamiya}}, \bibinfo {author} {\bibfnamefont {H.}~\bibnamefont {Zhang}}, \bibinfo {author} {\bibfnamefont {S.}~\bibnamefont {Lee}}, \bibinfo {author} {\bibfnamefont {A.~J.}\ \bibnamefont {Woods}}, \bibinfo {author} {\bibfnamefont {M.~O.}\ \bibnamefont {Ajeesh}}, \bibinfo {author} {\bibfnamefont {M.~G.}\ \bibnamefont {Gonzalez}}, \bibinfo {author} {\bibfnamefont {B.}~\bibnamefont {Bernu}}, \bibinfo {author} {\bibfnamefont {J.~W.}\ \bibnamefont {Villanova}}, \bibinfo {author} {\bibfnamefont {J.}~\bibnamefont {Xing}}, \bibinfo {author} {\bibfnamefont {Q.}~\bibnamefont {Huang}}, \bibinfo {author} {\bibfnamefont {Q.}~\bibnamefont {Zhang}}, \bibinfo {author} {\bibfnamefont {J.}~\bibnamefont {Ma}}, \bibinfo {author} {\bibfnamefont {E.~S.}\ \bibnamefont {Choi}}, \bibinfo {author} {\bibfnamefont {D.~M.}\ \bibnamefont {Pajerowski}}, \bibinfo {author} {\bibfnamefont {H.}~\bibnamefont
  {Zhou}}, \bibinfo {author} {\bibfnamefont {A.~S.}\ \bibnamefont {Sefat}}, \bibinfo {author} {\bibfnamefont {S.}~\bibnamefont {Okamoto}}, \bibinfo {author} {\bibfnamefont {T.}~\bibnamefont {Berlijn}}, \bibinfo {author} {\bibfnamefont {L.}~\bibnamefont {Messio}}, \bibinfo {author} {\bibfnamefont {R.}~\bibnamefont {Movshovich}}, \bibinfo {author} {\bibfnamefont {C.~D.}\ \bibnamefont {Batista}},\ and\ \bibinfo {author} {\bibfnamefont {D.~A.}\ \bibnamefont {Tennant}},\ }\bibfield  {title} {\bibinfo {title} {Nonlinear magnons and exchange hamiltonians of the delafossite proximate quantum spin liquid candidates {${\text{KYbSe}}_{2}$ and ${\text{NaYbSe}}_{2}$}},\ }\href {https://doi.org/10.1103/PhysRevB.109.014425} {\bibfield  {journal} {\bibinfo  {journal} {Phys. Rev. B}\ }\textbf {\bibinfo {volume} {109}},\ \bibinfo {pages} {014425} (\bibinfo {year} {2024}{\natexlab{b}})}\BibitemShut {NoStop}%
\bibitem [{\citenamefont {Zhu}\ and\ \citenamefont {White}(2015)}]{PhysRevB.92.041105}%
  \BibitemOpen
  \bibfield  {author} {\bibinfo {author} {\bibfnamefont {Z.}~\bibnamefont {Zhu}}\ and\ \bibinfo {author} {\bibfnamefont {S.~R.}\ \bibnamefont {White}},\ }\bibfield  {title} {\bibinfo {title} {Spin liquid phase of the $s=\frac{1}{2}\phantom{\rule{4.pt}{0ex}}{J}_{1}\ensuremath{-}{J}_{2}$ heisenberg model on the triangular lattice},\ }\href {https://doi.org/10.1103/PhysRevB.92.041105} {\bibfield  {journal} {\bibinfo  {journal} {Phys. Rev. B}\ }\textbf {\bibinfo {volume} {92}},\ \bibinfo {pages} {041105} (\bibinfo {year} {2015})}\BibitemShut {NoStop}%
\bibitem [{\citenamefont {Hu}\ \emph {et~al.}(2015)\citenamefont {Hu}, \citenamefont {Gong}, \citenamefont {Zhu},\ and\ \citenamefont {Sheng}}]{PhysRevB.92.140403}%
  \BibitemOpen
  \bibfield  {author} {\bibinfo {author} {\bibfnamefont {W.-J.}\ \bibnamefont {Hu}}, \bibinfo {author} {\bibfnamefont {S.-S.}\ \bibnamefont {Gong}}, \bibinfo {author} {\bibfnamefont {W.}~\bibnamefont {Zhu}},\ and\ \bibinfo {author} {\bibfnamefont {D.~N.}\ \bibnamefont {Sheng}},\ }\bibfield  {title} {\bibinfo {title} {Competing spin-liquid states in the spin-$\frac{1}{2}$ heisenberg model on the triangular lattice},\ }\href {https://doi.org/10.1103/PhysRevB.92.140403} {\bibfield  {journal} {\bibinfo  {journal} {Phys. Rev. B}\ }\textbf {\bibinfo {volume} {92}},\ \bibinfo {pages} {140403} (\bibinfo {year} {2015})}\BibitemShut {NoStop}%
\bibitem [{\citenamefont {Iqbal}\ \emph {et~al.}(2016)\citenamefont {Iqbal}, \citenamefont {Hu}, \citenamefont {Thomale}, \citenamefont {Poilblanc},\ and\ \citenamefont {Becca}}]{PhysRevB.93.144411}%
  \BibitemOpen
  \bibfield  {author} {\bibinfo {author} {\bibfnamefont {Y.}~\bibnamefont {Iqbal}}, \bibinfo {author} {\bibfnamefont {W.-J.}\ \bibnamefont {Hu}}, \bibinfo {author} {\bibfnamefont {R.}~\bibnamefont {Thomale}}, \bibinfo {author} {\bibfnamefont {D.}~\bibnamefont {Poilblanc}},\ and\ \bibinfo {author} {\bibfnamefont {F.}~\bibnamefont {Becca}},\ }\bibfield  {title} {\bibinfo {title} {Spin liquid nature in the heisenberg ${J}_{1}\ensuremath{-}{J}_{2}$ triangular antiferromagnet},\ }\href {https://doi.org/10.1103/PhysRevB.93.144411} {\bibfield  {journal} {\bibinfo  {journal} {Phys. Rev. B}\ }\textbf {\bibinfo {volume} {93}},\ \bibinfo {pages} {144411} (\bibinfo {year} {2016})}\BibitemShut {NoStop}%
\bibitem [{\citenamefont {Saadatmand}\ and\ \citenamefont {McCulloch}(2016)}]{PhysRevB.94.121111}%
  \BibitemOpen
  \bibfield  {author} {\bibinfo {author} {\bibfnamefont {S.~N.}\ \bibnamefont {Saadatmand}}\ and\ \bibinfo {author} {\bibfnamefont {I.~P.}\ \bibnamefont {McCulloch}},\ }\bibfield  {title} {\bibinfo {title} {Symmetry fractionalization in the topological phase of the spin-$\frac{1}{2}$ ${J}_{1}\text{\ensuremath{-}}{J}_{2}$ triangular heisenberg model},\ }\href {https://doi.org/10.1103/PhysRevB.94.121111} {\bibfield  {journal} {\bibinfo  {journal} {Phys. Rev. B}\ }\textbf {\bibinfo {volume} {94}},\ \bibinfo {pages} {121111} (\bibinfo {year} {2016})}\BibitemShut {NoStop}%
\bibitem [{\citenamefont {Wietek}\ and\ \citenamefont {L\"auchli}(2017)}]{PhysRevB.95.035141}%
  \BibitemOpen
  \bibfield  {author} {\bibinfo {author} {\bibfnamefont {A.}~\bibnamefont {Wietek}}\ and\ \bibinfo {author} {\bibfnamefont {A.~M.}\ \bibnamefont {L\"auchli}},\ }\bibfield  {title} {\bibinfo {title} {Chiral spin liquid and quantum criticality in extended $s=\frac{1}{2}$ heisenberg models on the triangular lattice},\ }\href {https://doi.org/10.1103/PhysRevB.95.035141} {\bibfield  {journal} {\bibinfo  {journal} {Phys. Rev. B}\ }\textbf {\bibinfo {volume} {95}},\ \bibinfo {pages} {035141} (\bibinfo {year} {2017})}\BibitemShut {NoStop}%
\bibitem [{\citenamefont {Gong}\ \emph {et~al.}(2017)\citenamefont {Gong}, \citenamefont {Zhu}, \citenamefont {Zhu}, \citenamefont {Sheng},\ and\ \citenamefont {Yang}}]{PhysRevB.96.075116}%
  \BibitemOpen
  \bibfield  {author} {\bibinfo {author} {\bibfnamefont {S.-S.}\ \bibnamefont {Gong}}, \bibinfo {author} {\bibfnamefont {W.}~\bibnamefont {Zhu}}, \bibinfo {author} {\bibfnamefont {J.-X.}\ \bibnamefont {Zhu}}, \bibinfo {author} {\bibfnamefont {D.~N.}\ \bibnamefont {Sheng}},\ and\ \bibinfo {author} {\bibfnamefont {K.}~\bibnamefont {Yang}},\ }\bibfield  {title} {\bibinfo {title} {Global phase diagram and quantum spin liquids in a spin-$\frac{1}{2}$ triangular antiferromagnet},\ }\href {https://doi.org/10.1103/PhysRevB.96.075116} {\bibfield  {journal} {\bibinfo  {journal} {Phys. Rev. B}\ }\textbf {\bibinfo {volume} {96}},\ \bibinfo {pages} {075116} (\bibinfo {year} {2017})}\BibitemShut {NoStop}%
\bibitem [{\citenamefont {Hu}\ \emph {et~al.}(2019)\citenamefont {Hu}, \citenamefont {Zhu}, \citenamefont {Eggert},\ and\ \citenamefont {He}}]{PhysRevLett.123.207203}%
  \BibitemOpen
  \bibfield  {author} {\bibinfo {author} {\bibfnamefont {S.}~\bibnamefont {Hu}}, \bibinfo {author} {\bibfnamefont {W.}~\bibnamefont {Zhu}}, \bibinfo {author} {\bibfnamefont {S.}~\bibnamefont {Eggert}},\ and\ \bibinfo {author} {\bibfnamefont {Y.-C.}\ \bibnamefont {He}},\ }\bibfield  {title} {\bibinfo {title} {Dirac spin liquid on the spin-$1/2$ triangular heisenberg antiferromagnet},\ }\href {https://doi.org/10.1103/PhysRevLett.123.207203} {\bibfield  {journal} {\bibinfo  {journal} {Phys. Rev. Lett.}\ }\textbf {\bibinfo {volume} {123}},\ \bibinfo {pages} {207203} (\bibinfo {year} {2019})}\BibitemShut {NoStop}%
\bibitem [{\citenamefont {Daszkiewicz}\ \emph {et~al.}(2008)\citenamefont {Daszkiewicz}, \citenamefont {Gulay}, \citenamefont {Shemet},\ and\ \citenamefont {Pietraszko}}]{Daszkiewicz2008}%
  \BibitemOpen
  \bibfield  {author} {\bibinfo {author} {\bibfnamefont {M.}~\bibnamefont {Daszkiewicz}}, \bibinfo {author} {\bibfnamefont {L.~D.}\ \bibnamefont {Gulay}}, \bibinfo {author} {\bibfnamefont {V.~Y.}\ \bibnamefont {Shemet}},\ and\ \bibinfo {author} {\bibfnamefont {A.}~\bibnamefont {Pietraszko}},\ }\bibfield  {title} {\bibinfo {title} {Comparative investigation of the crystal structure of {LnCuSe$_2$} compounds ({Ln = Tb, Dy, Ho, Er, Tm, Yb and Lu})},\ }\href {https://doi.org/https://doi.org/10.1002/zaac.200800039} {\bibfield  {journal} {\bibinfo  {journal} {Zeitschrift für anorganische und allgemeine Chemie}\ }\textbf {\bibinfo {volume} {634}},\ \bibinfo {pages} {1201} (\bibinfo {year} {2008})}\BibitemShut {NoStop}%
\bibitem [{\citenamefont {Ranjith}\ \emph {et~al.}(2019)\citenamefont {Ranjith}, \citenamefont {Luther}, \citenamefont {Reimann}, \citenamefont {Schmidt}, \citenamefont {Schlender}, \citenamefont {Sichelschmidt}, \citenamefont {Yasuoka}, \citenamefont {Strydom}, \citenamefont {Skourski}, \citenamefont {Wosnitza}, \citenamefont {K\"uhne}, \citenamefont {Doert},\ and\ \citenamefont {Baenitz}}]{PhysRevB.100.224417}%
  \BibitemOpen
  \bibfield  {author} {\bibinfo {author} {\bibfnamefont {K.~M.}\ \bibnamefont {Ranjith}}, \bibinfo {author} {\bibfnamefont {S.}~\bibnamefont {Luther}}, \bibinfo {author} {\bibfnamefont {T.}~\bibnamefont {Reimann}}, \bibinfo {author} {\bibfnamefont {B.}~\bibnamefont {Schmidt}}, \bibinfo {author} {\bibfnamefont {P.}~\bibnamefont {Schlender}}, \bibinfo {author} {\bibfnamefont {J.}~\bibnamefont {Sichelschmidt}}, \bibinfo {author} {\bibfnamefont {H.}~\bibnamefont {Yasuoka}}, \bibinfo {author} {\bibfnamefont {A.~M.}\ \bibnamefont {Strydom}}, \bibinfo {author} {\bibfnamefont {Y.}~\bibnamefont {Skourski}}, \bibinfo {author} {\bibfnamefont {J.}~\bibnamefont {Wosnitza}}, \bibinfo {author} {\bibfnamefont {H.}~\bibnamefont {K\"uhne}}, \bibinfo {author} {\bibfnamefont {T.}~\bibnamefont {Doert}},\ and\ \bibinfo {author} {\bibfnamefont {M.}~\bibnamefont {Baenitz}},\ }\bibfield  {title} {\bibinfo {title} {Anisotropic field-induced ordering in the triangular-lattice quantum spin liquid {${\mathrm{NaYbSe}}_{2}$}},\ }\href
  {https://doi.org/10.1103/PhysRevB.100.224417} {\bibfield  {journal} {\bibinfo  {journal} {Phys. Rev. B}\ }\textbf {\bibinfo {volume} {100}},\ \bibinfo {pages} {224417} (\bibinfo {year} {2019})}\BibitemShut {NoStop}%
\bibitem [{\citenamefont {Xing}\ \emph {et~al.}(2021)\citenamefont {Xing}, \citenamefont {Sanjeewa}, \citenamefont {May},\ and\ \citenamefont {Sefat}}]{10.1063/5.0071161}%
  \BibitemOpen
  \bibfield  {author} {\bibinfo {author} {\bibfnamefont {J.}~\bibnamefont {Xing}}, \bibinfo {author} {\bibfnamefont {L.~D.}\ \bibnamefont {Sanjeewa}}, \bibinfo {author} {\bibfnamefont {A.~F.}\ \bibnamefont {May}},\ and\ \bibinfo {author} {\bibfnamefont {A.~S.}\ \bibnamefont {Sefat}},\ }\bibfield  {title} {\bibinfo {title} {Synthesis and anisotropic magnetism in quantum spin liquid candidates {AYbSe2 (A = K and Rb)}},\ }\href {https://doi.org/10.1063/5.0071161} {\bibfield  {journal} {\bibinfo  {journal} {APL Materials}\ }\textbf {\bibinfo {volume} {9}},\ \bibinfo {pages} {111104} (\bibinfo {year} {2021})}\BibitemShut {NoStop}%
\bibitem [{\citenamefont {Xing}\ \emph {et~al.}(2019)\citenamefont {Xing}, \citenamefont {Sanjeewa}, \citenamefont {Kim}, \citenamefont {Stewart}, \citenamefont {Podlesnyak},\ and\ \citenamefont {Sefat}}]{PhysRevB.100.220407}%
  \BibitemOpen
  \bibfield  {author} {\bibinfo {author} {\bibfnamefont {J.}~\bibnamefont {Xing}}, \bibinfo {author} {\bibfnamefont {L.~D.}\ \bibnamefont {Sanjeewa}}, \bibinfo {author} {\bibfnamefont {J.}~\bibnamefont {Kim}}, \bibinfo {author} {\bibfnamefont {G.~R.}\ \bibnamefont {Stewart}}, \bibinfo {author} {\bibfnamefont {A.}~\bibnamefont {Podlesnyak}},\ and\ \bibinfo {author} {\bibfnamefont {A.~S.}\ \bibnamefont {Sefat}},\ }\bibfield  {title} {\bibinfo {title} {Field-induced magnetic transition and spin fluctuations in the quantum spin-liquid candidate {${\mathrm{CsYbSe}}_{2}$}},\ }\href {https://doi.org/10.1103/PhysRevB.100.220407} {\bibfield  {journal} {\bibinfo  {journal} {Phys. Rev. B}\ }\textbf {\bibinfo {volume} {100}},\ \bibinfo {pages} {220407} (\bibinfo {year} {2019})}\BibitemShut {NoStop}%
\bibitem [{\citenamefont {Kittel}(2004)}]{Kittel2004}%
  \BibitemOpen
  \bibfield  {author} {\bibinfo {author} {\bibfnamefont {C.}~\bibnamefont {Kittel}},\ }\href {http://www.amazon.com/Introduction-Solid-Physics-Charles-Kittel/dp/047141526X/ref=dp_ob_title_bk} {\emph {\bibinfo {title} {{Introduction to Solid State Physics}}}}\ (\bibinfo  {publisher} {John Wiley \& Sons, Ltd},\ \bibinfo {year} {2004})\BibitemShut {NoStop}%
\end{thebibliography}%

\end{document}